\documentclass[12pt]{article}
\pdfoutput=1
\usepackage{amssymb,amsfonts}
\usepackage{graphicx}
\begin{document}
\baselineskip 18pt

\title{Low-temperature asymptotic of the transverse dynamical structure factor for a magnetically polarized $XX$ chain}
\author{P.N. Bibikov}
\date{\it Russian State Hydrometeorological University, Saint-Petersburg, Russia}

\maketitle

\vskip5mm

\begin{abstract}
Dyson equation for the real two-time commutator retarded one-magnon Green function of the ferromagnetically polarized XX chain is suggested following the Plakida-Tserkovnikov algorithm.
Starting from this result a low-temperature integral representation for the corresponding magnon self energy is obtained by the truncated form factor expansion however without any resummations.
Within the suggested approach the low-temperature asymptotics of the transverse dynamical structure factor may be readily studied. Some obtained line shapes are presented.
\end{abstract}

\maketitle

\section{Introduction}

Dynamical structure factor of a magnetic compound is one of its most important characteristics directly measurable by neutron scattering \cite{1}.
The corresponding theoretical investigations on this direction are now far from completeness even for low-dimensional spin models \cite{2}. The simplest of them is the 1D XX chain related to the Hamiltonian
\begin{equation}
\hat H=-\sum_{n=1}^N\Big[\frac{J}{2}\Big({\bf S}^+_n{\bf S}^-_{n+1}+{\bf S}^-_n{\bf S}^+_{n+1}\Big)+h\Big({\bf S}_n^z-\frac{1}{2}\Big)\Big],\qquad{\bf S}_{N+1}\equiv{\bf S}_1.
\end{equation}
Here ${\bf S}_n^z$ and ${\bf S}_n^{\pm}={\bf S}_n^x\pm i{\bf S}_n^y$ is the standard triple of spin-1/2 operators acting in the corresponding copy of the space ${\mathbb C}^2$ associated with $n$-th site.

In the gapped (massive) regime \cite{3} related to the condition
\begin{equation}
E_{gap}(h)=h-|J|>0,
\end{equation}
the Hilbert space ${\cal H}$ of the model (which is tensor product of $N$ copies of ${\mathbb C}^2$) splits on the direct sum of $m$-magnon sectors
\begin{equation}
{\cal H}=\oplus_{m=0}^N{\cal H}_m,\qquad\Big(\frac{N}{2}-\sum_{n=1}^N{\bf S}_n^z\Big)\Big|_{{\cal H}_m}=m,
\end{equation}
where the one-dimensional sector ${\cal H}_0$ is generated by the ferromagnetically polarized zero energy ground state
\begin{equation}
|\emptyset\rangle=|\uparrow\rangle_1\otimes\dots\otimes|\uparrow\rangle_N.
\end{equation}
Here $|\uparrow\rangle_n$ and $|\downarrow\rangle_n$ are the spin polarized local states corresponding to $n$-th site.

The corresponding transverse dynamical structure factor (TDSF) is alternatively defined by one of the formulas (as usual $\beta\equiv1/(k_BT)$)
\begin{eqnarray}
&&g(t,n,T)=\lim_{N\rightarrow\infty}\frac{1}{Z(T,N)}{\rm Tr}\Big({\rm e}^{-\beta\hat H}
{\bf S}_n^+(t){\bf S}_0^-\Big),\qquad{\bf S}_n^+(t)\equiv{\rm e}^{i\hat H t}{\bf S}_n^+{\rm e}^{-i\hat H t},\\
&&S(\omega,q,T)=\lim_{N\rightarrow\infty}\frac{1}{Z(T,N)}\sum_{\mu,\nu}{\rm e}^{-\beta E_{\nu}}
|\langle\nu|{\bf S}^+(q)|\mu\rangle|^2\delta(\omega+E_{\nu}-E_{\mu}),
\end{eqnarray}
related to the space-time and spectral representations. Here $Z(T,N)$ is the partition function, the two parameters $\mu$ and $\nu$ enumerate an eigenbasis of ${\hat H}$ and
\begin{equation}
{\bf S}^{\pm}(q)\equiv\frac{1}{\sqrt{N}}\sum_{n=1}^N{\rm e}^{-iqn}{\bf S}^{\pm}_n.
\end{equation}
In (7) it is implied that
\begin{equation}
-\pi<q\leq\pi,\qquad{\rm e}^{iqN}=1.
\end{equation}
The equivalence between (5) and (6) is expressed by the well known relation
\begin{equation}
g(t,n,T)=\frac{1}{2\pi}\int_{-\infty}^{\infty}d\omega\int_{-\pi}^{\pi}dq{\rm e}^{i(qn-\omega t)}S(\omega,q,T),
\end{equation}
(proved in the Appendix A). The definition (5) is more compact however just the function $S(\omega,q,T)$ is measurable in neutron scattering
experiments \cite{1}.

Although an exact formula for the longitudinal dynamic structure factor (${\bf S}_n^+(t){\bf S}_0^-\longrightarrow{\bf S}_n^z(t){\bf S}_0^z$ in (5) or ${\bf S}^+(q)\longrightarrow{\bf S}^z(q)$ in (6))
was obtained long ago by various approaches (see Refs. in \cite{3}) the corresponding result for the TDSF at present time is lack. The first essential progress in this direction
was achieved in \cite{4} where the large-time asymptotic for the function $g(t,n,T)$ was derived by a combination of approaches developed
previously or classical and quantum integrable systems \cite{5,6}. Recently \cite{7} the problem was attacked again in the framework of the
Quantum Transfer Matrix approach \cite{8}. Contrary to \cite{4} where an asymptotic formula for $g(t,n,T)$ was obtained analytically the authors of \cite{7} employed on the final stage
purely numerical methods.

Since both the approaches \cite{4} and \cite{7} are based on the machinery of integrable systems \cite{5,6,8} they operate with the {\it total spectrum} of (1) and as a result give predictions valid in the {\it whole
diapason} of temperatures. It is a common opinion \cite{9,10,11,12} however that all the properties of a {\it gapped} system in the {\it low-temperature} asymptotic regime
\begin{equation}
{\rm e}^{-\beta E_{gap}}\ll1,
\end{equation}
depend only on its {\it few-particle spectrum}.
The corresponding machinery for evaluation of the low-temperature asymptotics for statical physical quantities (free energy density and its derivatives) as series expansions governed by ${\rm e}^{-\beta E_{gap}}$
is well developed \cite{9,10,11,12}. However a direct transfer of these methods on TDSF results in a problem. Really if one suggest the straightforward low temperature expansion
\begin{equation}
S(\omega,q,T)=\sum_{m=0}^{\infty}S_m(\omega,q,T),\qquad S_m(\omega,q,T)=O\Big({\rm e}^{-m\beta E_{gap}}\Big),
\end{equation}
for TDSF, then according to the spectral representation (6) $S_0(\omega,q,T)$ does not depend on $T$ and (do not forget that $Z(0,N)=1$) has the form
\begin{equation}
S_0(\omega,q)=S(\omega,q,0)=\lim_{N\rightarrow\infty}\sum_k|\langle\emptyset|{\bf S}^+(q)|k\rangle|^2\delta(\omega-E_{magn}(k)),
\end{equation}
where
\begin{equation}
|k\rangle=\frac{1}{\sqrt{N}}\sum_{n=1}^N{\rm e}^{ikn}{\bf S}_n^-|\emptyset\rangle={\bf S}^-(-k)|\emptyset\rangle,\qquad{\rm e}^{ikN}=1,
\end{equation}
is a normalized one magnon state
\begin{equation}
\langle\tilde k|k\rangle=\frac{1}{N}\sum_{n=1}^N{\rm e}^{i(k-\tilde k)n}=\delta_{k\tilde k},
\end{equation}
with energy \cite{3}
\begin{equation}
E_{magn}(k)=h-J\cos{k}=h-|J|\cos{(k-k_{gap})},\qquad k_{gap}=\left\{\begin{array}{rcl}
0,\quad J>0,\\
\pi,\quad J<0.
\end{array}\right.
\end{equation}
A simple calculation gives the singular result
\begin{equation}
S_0(\omega,q)=\delta(\omega-E_{magn}(q)),
\end{equation}
which {\it can not be repaired} by any finite number of higher order terms. At the same time it is a common opinion that the finite temperature line shape of TDSF should
be smooth \cite{1}.

A modified approach for evaluation of $S(\omega,q,T)$ at nonzero temperatures was suggested in \cite{13,14,15} according to the well known formula \cite{1}
\begin{equation}
(1-{\rm e}^{-\beta\omega})S(\omega,q,T)=-\frac{1}{\pi}{\rm Im}\chi(\omega,q,T),
\end{equation}
which at $\omega\neq0$ is equivalent to
\begin{equation}
S(\omega,q,T)=-\frac{1}{\pi(1-{\rm e}^{-\beta\omega})}{\rm Im}\chi(\omega,q,T),\qquad\omega\neq0.
\end{equation}
Here $\chi(\omega,q,T)$ is the dynamical magnetic susceptibility
and at the same time the real two-time commutator retarded one-magnon Green function \cite{16}
\begin{equation}
\chi(\omega,q,T)=\lim_{N\rightarrow\infty}\langle\langle{\bf S}^+(q),{\bf S}^-(-q)\rangle\rangle_{\omega},
\end{equation}
where for two operators $A$ and $B$ there are two equivalent representations of $\langle\langle A,B\rangle\rangle_{\omega}$
\begin{eqnarray}
&&\langle\langle A,B\rangle\rangle_{\omega}\equiv\frac{1}{i}\int_0^{\infty}dt{\rm e}^{i(\omega+i\epsilon)t}\langle[A(t),B]\rangle,\\
&&\langle\langle A,B\rangle\rangle_{\omega} \equiv\frac{1}{i}\int_0^{\infty}dt{\rm e}^{i(\omega+i\epsilon)t}\langle[A,B(-t)]\rangle.
\end{eqnarray}
As usual
\begin{equation}
\langle A\rangle\equiv\frac{1}{Z(T,N)}{\rm tr}\Big({\rm e}^{-\beta\hat H}A\Big),\qquad A(t)\equiv{\rm e}^{iHt}A{\rm e}^{-iHt}.
\end{equation}

Of course direct use of (19) can not repair the singular result (16). Really using the well known spectral decomposition (reproved in Appendix A)
\begin{equation}
\langle\langle{\bf S}^+(q),{\bf S}^-(-q)\rangle\rangle_{\omega}=\frac{1}{Z(T,N)}\sum_{\mu,\nu}\frac{{\rm e}^{-\beta E_{\nu}}-{\rm e}^{-\beta E_{\mu}}}{\omega+E_{\nu}-E_{\mu}+i\epsilon}
|\langle\nu|{\bf S}^+(q)|\mu\rangle|^2,
\end{equation}
one readily gets from (7) and (13) the zero temperature expression
\begin{equation}
\chi(\omega,q,0)=\lim_{N\rightarrow\infty}\sum_k\frac{|\langle\emptyset|{\bf S}^+(q)|k\rangle|^2}{\omega-E_{magn}(k)+i\epsilon}
=\frac{1}{\omega-E_{magn}(q)+i\epsilon},
\end{equation}
which according to the well known formula
\begin{equation}
{\rm Im}\frac{1}{x+i\epsilon}=-\pi\delta(x),
\end{equation}
directly gives (16).

It was however suggested in \cite{13,14,15} that at $T>0$ the Green function (19) should satisfy the Dyson equation and hence may be represented in the form
\begin{equation}
\chi(\omega,q,T)=\frac{1}{\omega-E_{magn}(q)-\Sigma(\omega,q,T)},
\end{equation}
where $\Sigma(\omega,q,T)$ is the so called self-energy. If now the equation
\begin{equation}
\omega-E_{magn}(q)-\Sigma(\omega,q,T)=0,
\end{equation}
has no solutions for {\it real} $\omega$ and $q$ then the right side of (26) is regular for real $\omega$ and $q$.
For example the singularity removes if for real $\omega$ and $q$
\begin{equation}
{\rm Im}\Sigma(\omega,q,T)\neq0.
\end{equation}

So in order to obtain a smooth expression for TDSF at $\omega\neq0$ it is necessary to turn from (11) to an alternative expansion for the self energy
\begin{equation}
\Sigma(\omega,q,T)=\sum_{m=1}^{\infty}\Sigma_m(\omega,q,T),\qquad\Sigma_m(\omega,q,T)=O({\rm e}^{-m\beta E_{gap}}).
\end{equation}
But the source of the low temperature expansion in \cite{13,14,15} still remains the spectral decomposition (23) for $\chi(\omega,q,T)$
not for $\Sigma(\omega,q,T)$. That is why a passage from (29) to (11) may be realized only by the resummation procedure. Namely in zero order (24) gives
$\chi_0(\omega,q)=\chi(\omega,q,0)$. Hence the first order formula
\begin{equation}
\frac{1}{\omega-E_{magn}(q)+i\epsilon}+\chi_1(\omega,q,T)+\dots=\frac{1}{\omega-E_{magn}(q)-\Sigma_1(\omega,q,T)+\dots},
\end{equation}
directly yields
\begin{equation}
\Sigma_1(\omega,q,T)=(\omega-E_{magn}(q))^2\chi_1(\omega,q,T).
\end{equation}
From (31) follows that in order to obtain $\Sigma_1(\omega,q,T)$ we need to know $\chi_1(\omega,q,T)$ and so on. It may be readily seen however that even an evaluation of $\chi_1(\omega,q,T)$
is a rather cumbersome problem. A special question is a rigorous proof of the representation (26). To the author knowledge structure of the Matsubara temperature Green functions used in \cite{13,14,15} may be studied
only {\it perturbatively} according to a detailed analysis of Feynman diagrams. The latter procedure is rather straightforward for Bose and Fermi systems but becomes
complicated for spin ones where the operator algebra is more complex. In fact the correct form of the temperature spin Green function (for which in \cite{13,14,15}
was postulated the representation (26)) is not yet completely  established \cite{17,18}.

At the same time for the real two-time Green function (19) the representation (26) may be proved {\it analytically} within the approach suggested and developed
by N. M. Plakida and Yu. A. Tserkovnikov \cite{19,20,21,22}. Moreover as it is shown in the paper an evaluation of $\Sigma_1(\omega,q,T)$ in this framework is rather simple and does not need
a preliminary knowledge of $\chi_1(\omega,q,T)$ (so that the resummation does not occur).

The paper is organized as follows. In Sect. 2 we represent the two-magnon sector of the model \cite{3} in the form which seems more convenient for the further calculations. In Sect. 3
applying the Plakida-Tserkovnikov approach to the model (1) we obtain the Dyson equation and the form factor representation for the self energy $\Sigma(\omega,q,T)$. In Sect. 4 using the truncated form factor expansion
we calculate $\Sigma_1(\omega,q,T)$ (the first term in (29)). In Sections 5,6 and 7 for the special values  $q=0,\pi,\pi/2$ we reduce the general expression for
$\Sigma_1(\omega,q,T)$ to forms more convenient for numerical calculations. We also present some examples of line shapes obtained with a use of MATLAB. Finitely in Sect. 8 we summarize the obtained results
and point some aspects which were not elucidated.

\section{The two-magnon excitations}

A two-magnon state has the form
\begin{equation}
|2-magn\rangle=\sum_{n_1<n_2}\psi_{n_1,n_2}{\bf S}^-_{n_1}{\bf S}^-_{n_2}|\emptyset\rangle,
\end{equation}
where the wave function satisfies the ${\rm Schr\ddot odinger}$ equation
\begin{eqnarray}
&&2h\psi_{n_1,n_2}-\frac{J}{2}\Big(\psi_{n_1-1,n_2}+\psi_{n_1+1,n_2}+\psi_{n_1,n_2-1}+\psi_{n_1,n_2+1}\Big)=E\psi_{n_1,n_2},\qquad n_2-n_1>1,\nonumber\\
&&2h\psi_{n,n+1}-\frac{J}{2}\Big(\psi_{n-1,n+1}+\psi_{n,n+2}\Big)=E\psi_{n,n+1}.
\end{eqnarray}
We also suggest the periodicity and normalization conditions
\begin{equation}
\psi_{n_2,n_1+N}=\psi_{n_1,n_2},\qquad\sum_{1\leq n_1<n_2\leq N}|\psi_{n_1,n_2}|^2=1.
\end{equation}
It is well known \cite{3,6} that (33), (34) yield only {\it scattering} two-magnon states. The corresponding eigenbasis has the form
\begin{equation}
|k,\kappa\rangle=\frac{2}{N}\sum_{n_1<n_2}{\rm e}^{ik(n_1+n_2)/2}\sin{\kappa(n_2-n_1)}{\bf S}^-_{n_1}{\bf S}^-_{n_2}|\emptyset\rangle.
\end{equation}
The related energies are
\begin{equation}
E_{scatt}(k,\kappa)=E_{magn}(k/2-\kappa)+E_{magn}(k/2+\kappa)=2(h-J\cos{k/2}\cos{\kappa}).
\end{equation}
According to the periodicity condition in (34)
\begin{equation}
{\rm e}^{ikN}=1,\qquad0<\kappa<\pi,\qquad{\rm e}^{i(k/2+\kappa)N}=-1.
\end{equation}
The normalization condition in (34) takes the form
\begin{equation}
\langle k,\kappa|\tilde k,\tilde\kappa\rangle=\delta_{k,\tilde k}\delta_{\kappa,\tilde\kappa}.
\end{equation}
Implying
\begin{equation}
-\pi<k\leq\pi\Longrightarrow\cos{\frac{k}{2}}\geq0,
\end{equation}
one readily gets from (36)
\begin{equation}
E_{down}(k)\leq E_{scatt}(k,\kappa)\leq E_{up}(k),
\end{equation}
where the down and up boundaries of the two-magnon scattering zone are
\begin{equation}
E_{down}(k)=2h-2|J|\cos{\frac{k}{2}},\qquad E_{up}(k)=2h+2|J|\cos{\frac{k}{2}}.
\end{equation}

\section{Dyson equation and self-energy}

From (20) and (21) follow the equations of motion
\begin{eqnarray}
(\omega+i\epsilon)\langle\langle A,B\rangle\rangle_{\omega}=\langle[A,B]\rangle_{N}
+\langle\langle[A,\hat H],B\rangle\rangle_{\omega},\\
(\omega+i\epsilon)\langle\langle A,B\rangle\rangle_{\omega}=\langle[A,B]\rangle_{N}
-\langle\langle A,[B,\hat H]\rangle\rangle_{\omega}.
\end{eqnarray}

Since
\begin{equation}
[{\bf S}^+(q),{\bf S}^-(-q)]=\frac{2}{N}\sum_{n=1}^N{\bf S}^z_n\equiv 2{\bf M}^z,
\end{equation}
one has from (42)
\begin{eqnarray}
&&(\omega+i\epsilon)\langle\langle{\bf S}^+(q),{\bf S}^-(-q)\rangle\rangle_{\omega}=\sigma(\beta)
+\langle\langle{\bf X}^+(q),{\bf S}^-(-q)\rangle\rangle_{\omega},\\
&&(\omega+i\epsilon)\langle\langle{\bf S}^+(q),{\bf S}^-(-q)\rangle\rangle_{\omega}=\sigma(\beta)
+\langle\langle{\bf S}^+(q),{\bf X}^-(-q)\rangle\rangle_{\omega},
\end{eqnarray}
where
\begin{equation}
\sigma(T)=2\langle{\bf M}^z\rangle=\langle[{\bf S}^+(q),{\bf S}^-(-q)]\rangle
\end{equation}
(according to the translation invariance $\sigma(\beta)=2\langle{\bf S}_0^z\rangle$) and
\begin{equation}
{\bf X}^{\pm}(q)\equiv\pm[{\bf S}^{\pm}(q),\hat H]
=h{\bf S}^{\pm}(q)-\frac{J}{\sqrt{N}}\sum_{n=1}^N{\rm e}^{-iqn}
\Big({\bf S}_{n-1}^{\pm}+{\bf S}_{n+1}^{\pm}\Big){\bf S}_n^z.
\end{equation}

Let us now extract from ${\bf X}^{\pm}(q)$ their {\it irreducible} (with respect to ${\bf S}^{\pm}(q)$) parts ${\bf Y}^{\pm}(q)$ taking
\begin{equation}
{\bf X}^{\pm}(q)=\theta_{\pm}(q,T){\bf S}^{\pm}(q)+{\bf Y}^{\pm}(q,T),
\end{equation}
and suggesting
\begin{equation}
\langle[{\bf Y}^+(q,T),{\bf S}^-(-q)]\rangle=\langle[{\bf S}^+(q),{\bf Y}^-(-q,T)]\rangle=0,
\end{equation}
or equivalently
\begin{equation}
\theta_+(q,T)=\frac{\langle[{\bf X}^+(q),{\bf S}^-(-q)]\rangle}{\langle[{\bf S}^+(q),{\bf S}^-(-q)]\rangle},\qquad
\theta_-(q,T)=\frac{\langle[{\bf S}^+(q),{\bf X}^-(-q)]\rangle}{\langle[{\bf S}^+(q),{\bf S}^-(-q)]\rangle}.
\end{equation}

Since
\begin{equation}
[{\bf X}^+(q),{\bf S}^-(-q)]=[{\bf S}^+(q),{\bf X}^-(-q)]=2h{\bf M}^z+V(q),
\end{equation}
where
\begin{eqnarray}
&&V(q)=\frac{J}{N}\sum_{n=1}^N\Big[\Big({\bf S}_{n-1}^-+{\bf S}_{n+1}^-\Big){\bf S}_n^+-4\cos{q}{\bf S}_n^z{\bf S}_{n+1}^z\Big]\nonumber\\
&&=\frac{2}{N}\Big[h\sum_{n=1}^N\Big(\frac{1}{2}-{\bf S}_n^z\Big)-\hat H\Big]-\frac{4J\cos{q}}{N}\sum_{n=1}^N{\bf S}_n^z{\bf S}_{n+1}^z,
\end{eqnarray}
one readily has from (44) and (51)-(53)
\begin{equation}
\theta(q,T)\equiv\theta_+(q,T)=\theta_-(q,T)=h+\frac{v(q,T)}{\sigma(T)},\qquad v(q,T)\equiv\langle V(q)\rangle.
\end{equation}
Substituting now (49) and (54) into (45) and (46) one readily pass from ${\bf X}^{\pm}(q)$ to their irreducible parts ${\bf Y}^{\pm}(q)$
\begin{eqnarray}
&&(\omega-\theta(q,T)+i\epsilon)\langle\langle{\bf S}^+(q),{\bf S}^-(-q)\rangle\rangle_{\omega}
=\sigma(T)+\langle\langle{\bf Y}^+(q,T),{\bf S}^-(-q)\rangle\rangle_{\omega},\\
&&(\omega-\theta(q,T)+i\epsilon)\langle\langle{\bf S}^+(q),{\bf S}^-(-q)\rangle\rangle_{\omega}
=\sigma(T)+\langle\langle{\bf S}^+(q,T),{\bf Y}^-(-q)\rangle\rangle_{\omega}.
\end{eqnarray}

Let us now apply (43) to $\langle\langle{\bf Y}^+(q,T),{\bf S}^-(-q)\rangle\rangle_{\omega}$. According to (48) and (50) one readily gets
\begin{equation}
(\omega+i\epsilon)\langle\langle{\bf Y}^+(q,T),{\bf S}^-(-q)\rangle\rangle_{\omega}=\langle\langle{\bf Y}^+(q,T),{\bf X}^-(-q)\rangle\rangle_{\omega},
\end{equation}
or after a substitution of (49) and (54)
\begin{equation}
(\omega-\theta(q,T)+i\epsilon)\langle\langle{\bf Y}^+(q,T),{\bf S}^-(-q)\rangle\rangle_{\omega}
=\langle\langle{\bf Y}^+(q,T),{\bf Y}^-(-q,T)\rangle\rangle_{\omega}.
\end{equation}
Expanding now the product
\begin{eqnarray}
&&\Big((\omega-\theta(q,T)+i\epsilon)\langle\langle{\bf S}^+(q),{\bf S}^-(-q)\rangle\rangle_{\omega}\Big)\langle\langle{\bf Y}^+(q,T),{\bf S}^-(-q)\rangle\rangle_{\omega}\nonumber\\
&&=\langle\langle{\bf S}^+(q),{\bf S}^-(-q)\rangle\rangle_{\omega}\Big((\omega-\theta(q,T)+i\epsilon)\langle\langle{\bf Y}^+(q,T),{\bf S}^-(-q)\rangle\rangle_{\omega}\Big),
\end{eqnarray}
in turn by (56) and (58) one gets
\begin{eqnarray}
&&(\sigma(T)+\langle\langle{\bf S}^+(q),{\bf Y}^-(-q,T)\rangle\rangle_{\omega})\langle\langle{\bf Y}^+(q,T),{\bf S}^-(-q)\rangle\rangle_{\omega}\nonumber\\
&&=\langle\langle{\bf S}^+(q),{\bf S}^-(-q)\rangle\rangle_{\omega}\langle\langle{\bf Y}^+(q,T),{\bf Y}^-(-q,T)\rangle\rangle_{\omega},
\end{eqnarray}
or equivalently
\begin{equation}
\sigma(T)\langle\langle{\bf Y}^+(q,T),{\bf S}^-(-q)\rangle\rangle_{\omega}
=\langle\langle{\bf S}^+(q),{\bf S}^-(-q)\rangle\rangle_{\omega}\langle\langle{\bf Y}^+(q,T),{\bf Y}^-(-q,T)\rangle\rangle^{(irr)}_{\omega},
\end{equation}
where for two operators $A$ and $B$
\begin{equation}
\langle\langle A,B\rangle\rangle^{(irr)}_{\omega}\equiv\langle\langle A,B\rangle\rangle_{\omega}
-\frac{\langle\langle A,{\bf S}^-(-q)\rangle\rangle_{\omega}\langle\langle{\bf S}^+(q),B\rangle\rangle_{\omega}}{\langle\langle{\bf S}^+(q),{\bf S}^-(-q)\rangle\rangle_{\omega}}.
\end{equation}
Now a substitution of (61) into (55) yields
\begin{equation}
\langle\langle{\bf S}^+(q),{\bf S}^-(-q)\rangle\rangle_{\omega}=G(\omega,q,T)
+G(\omega,q,T)\Pi(\omega,q,T)\langle\langle{\bf S}^+(q),{\bf S}^-(-q)\rangle\rangle_{\omega},
\end{equation}
or equivalently
\begin{equation}
\langle\langle{\bf S}^+(q),{\bf S}^-(-q)\rangle\rangle_{\omega}=\frac{1}{G^{-1}(\omega,q,T)-\Pi(\omega,q,T)},
\end{equation}
where
\begin{eqnarray}
&&G(\omega,q,T)\equiv\frac{\sigma(T)}{\omega-\theta(q,T)+i\epsilon},\\
&&\Pi(\omega,q,T)=\frac{1}{\sigma^2(T)}\langle\langle{\bf Y}^+(q,T),{\bf Y}^-(-q,T)\rangle\rangle^{(irr)}_{\omega}.
\end{eqnarray}

In \cite{19,20} (63) was associated with the Dyson equation. However $G(\omega,q,T)$ which plays in (63) a role of the "free" Green function does not coincide with (24). Moreover it depends on temperature
and using (47) and (51) may be represented in the form
\begin{equation}
G(\omega,q,T)=\frac{\displaystyle\langle[{\bf S}^+(q),{\bf S}^-(-q)]\rangle}
{\displaystyle\omega-\frac{\langle[[{\bf S}^+(q),\hat H],{\bf S}^-(-q)]\rangle}
{\langle[{\bf S}^+(q),{\bf S}^-(-q)]\rangle}+i\epsilon},
\end{equation}
associated with the Roth variational approximation \cite{23} for $\langle\langle{\bf S}^+(q),{\bf S}^-(-q)\rangle\rangle_{\omega}$. Nevertheless defining the self-energy
$\Sigma(\omega,q,T)$ according to the following relation
\begin{equation}
\omega-E_{magn}(q)-\Sigma(\omega,q,T)=G^{-1}(\omega,q,T)-\Pi(\omega,q,T),
\end{equation}
one readily turns from (64) to (26).

The low temperature expansion for $G^{-1}(\omega,q,T)$ may be readily obtained from (47), (54) and (65). Namely suggesting the asymptotic expansions
\begin{eqnarray}
&&\sigma(T)=\sigma_0+\sum_{m=1}^{\infty}\sigma_m(T),\qquad \sigma_m(T)=O\Big({\rm e}^{-\beta mE_{gap}}\Big)\nonumber\\
&&v(q,T)=v_0(q)+\sum_{m=1}^{\infty}v_m(q,T),\qquad v_m(q,T)=O\Big({\rm e}^{-\beta mE_{gap}}\Big),\nonumber\\
&&\theta(q,T)=\theta_0(q)+\sum_{m=1}^{\infty}\theta_m(q,T),\qquad \theta_m(q,T)=O\Big({\rm e}^{-\beta mE_{gap}}\Big),
\end{eqnarray}
and taking into account that according to (44) and (53)
\begin{equation}
2{\bf M}^z|\emptyset\rangle=|\emptyset\rangle,\qquad V|\emptyset\rangle=-J\cos{q}|\emptyset\rangle,
\end{equation}
one readily gets
\begin{equation}
\sigma_0=\langle\emptyset|2{\bf M}^z|\emptyset\rangle=1,\quad v_0(q)=\langle\emptyset|V|\emptyset\rangle=-J\cos{q},\quad\theta_0(q)=h+\frac{v_0(q)}{\sigma_0}=E_{magn}(q).
\end{equation}
Suggesting now an analogous low-temperature expansion
\begin{equation}
\Pi(\omega,q,T)=\sum_{m=1}^{\infty}\Pi_m(\omega,q,T),\qquad\Pi_m(\omega,q,T)=O\Big({\rm e}^{-m\beta E_{gap}}\Big),
\end{equation}
let us first prove that
\begin{equation}
\Pi_0(\omega,q,T)=0.
\end{equation}
Really according to definition (62) for arbitrary scalars $\lambda$ and $\mu$ one has
\begin{equation}
\langle\langle A,B\rangle\rangle^{(irr)}_{\omega}=\langle\langle A+\lambda{\bf S}^+(q),B+\mu{\bf S}^-(-q)\rangle\rangle^{(irr)}_{\omega}.
\end{equation}
This property allows to reduce (66) to a more convenient form
\begin{equation}
\Pi(\omega,q,T)=\frac{1}{\sigma^2(T)}\langle\langle{\bf Z}^+(q,T),{\bf Z}^-(-q,T)\rangle\rangle^{(irr)}_{\omega},
\end{equation}
where
\begin{equation}
\langle\emptyset|{\bf Z}^+(q,T)={\bf Z}^-(q,T)|\emptyset\rangle=0,
\end{equation}
so that (73)  becomes a consequence of the definition (62) and the spectral decomposition (A.5) applied to $\langle\langle{\bf Z}^+(q,T),{\bf Z}^-(-q,T)\rangle\rangle_{\omega}$,
$\langle\langle{\bf S}^+(q),{\bf Z}^-(-q,T)\rangle\rangle_{\omega}$ and $\langle\langle{\bf Z}^+(q,T),{\bf S}^-(-q)\rangle\rangle_{\omega}$.

Really (76) is obviously satisfied for the ($T$-independent) operator
\begin{eqnarray}
&&{\bf Z}^{\pm}(q)={\bf Y}^{\pm}(q,T)+(\theta(q,T)-E_{magn}(q)){\bf S}^{\pm}(q)\nonumber\\
&&=\frac{J}{\sqrt{N}}\sum_n{\rm e}^{-iqn}\Big({\bf S}_{n-1}^{\pm}
+{\bf S}_{n+1}^{\pm}\Big)\Big(\frac{1}{2}-{\bf S}_n^z\Big).
\end{eqnarray}
Hence (73) is proved.

Let us make now some estimations. According to (24), (71), (76) and the spectral decomposition (A.5) one has
\begin{eqnarray}
&&\sigma(T)=1+o({\rm e}^{-\beta E_{gap}}),\qquad\langle\langle{\bf S}^+(q),{\bf S}^-(-q)\rangle\rangle_{\omega}=O(1),\nonumber\\
&&\langle\langle{\bf Z}^+(q),{\bf S}^-(-q)\rangle\rangle_{\omega}=O({\rm e}^{-\beta E_{gap}}),
\qquad\langle\langle{\bf S}^+(q),{\bf Z}^-(-q)\rangle\rangle_{\omega}=O({\rm e}^{-\beta E_{gap}}).
\end{eqnarray}
Hence in the order $O({\rm e}^{-\beta E_{gap}})$ (75) reduces to
\begin{equation}
\Pi(\omega,q,T)=\langle\langle{\bf Z}^+(q),{\bf Z}^-(-q)\rangle\rangle_{\omega}+o({\rm e}^{-\beta E_{gap}}),
\end{equation}
(without any dependence on $\langle\langle{\bf S}^+(q),{\bf S}^-(-q)\rangle\rangle_{\omega}$ and hence {\it no resummation!}). Now (79) and the spectral decomposition (A.5) yield
\begin{equation}
\Pi_1(\omega,q,T)=\sum_k\sum_{\kappa}
\frac{{\rm e}^{-\beta E_{magn}(k-q)}|\langle k-q|{\bf Z}^+(q)|k,\kappa\rangle|^2}{\omega+E_{magn}(k-q)-E_{scatt}(k,\kappa)+i\epsilon}.
\end{equation}

\section{Truncated form factor expansion}

The matrix element in (80) may be readily calculated. Really according to (77)
\begin{equation}
{\bf Z}^+(q)\sum_{n_1<n_2}\psi_{n_1,n_2}{\bf S}^-_{n_1}{\bf S}^-_{n_2}|\emptyset\rangle=\frac{J}{\sqrt{N}}\sum_{n=1}^N{\rm e}^{-iqn}(\psi_{n-1,n}(k,\kappa)+\psi_{n,n+1}(k,\kappa)){\bf S}_n^-|\emptyset\rangle,
\end{equation}
so a substitution of (35) into (81) yields
\begin{equation}
{\bf Z}^+(q)|k,\kappa\rangle
=\frac{4J\sin{\kappa}}{N\sqrt{N}}\cos{\frac{k}{2}}\sum_n{\rm e}^{i(k-q)n}{\bf S}^-_n|\emptyset\rangle
=\frac{4J\sin{\kappa}}{N}\cos{\frac{k}{2}}|k-q\rangle.
\end{equation}
Hence
\begin{equation}
|\langle k-q|{\bf Z}^+(q)|k,\kappa\rangle|^2=\frac{16J^2\sin^2{\kappa}}{N^2}\cos^2{\frac{k}{2}}.
\end{equation}

Taking into account that $E_{scatt}(k,-\kappa)=E_{scatt}(k,\kappa)$ and using the $N\rightarrow\infty $ substitutions $\frac{1}{N}\sum_k\longrightarrow\frac{1}{2\pi}\int_{-\pi}^{\pi}dk$ and $\frac{1}{N}\sum_{\kappa}\longrightarrow\frac{1}{2\pi}\int_0^{\pi}d\kappa\longrightarrow\frac{1}{4\pi}\int_{-\pi}^{\pi}d\kappa$
one can obtain from (80) and (83) the $N=\infty$ expression
\begin{equation}
\Pi_1(\omega,q,T)=\frac{1}{\pi}\int_{-\pi}^{\pi}dk{\rm e}^{-\beta E_{magn}(k-q)}\Gamma(k,\omega,q),
\end{equation}
where
\begin{equation}
\Gamma(k,\omega,q)=\frac{2J^2}{\pi}\cos^2{\frac{k}{2}}\int_{-\pi}^{\pi}
\frac{\sin^2{\kappa}d\kappa}{\omega+E_{magn}(k-q)-E_{scatt}(k,\kappa)+i\epsilon}.
\end{equation}

Using the variables
\begin{equation}
z={\rm e}^{i\kappa},\quad a=J\cos{\frac{k}{2}},\quad b=\omega-h-J\cos{(k-q)}=\omega-h-|J|\cos{(k-q-k_{gap})},
\end{equation}
and taking into account (15) and (36) one may represent (85) in the form
\begin{equation}
\Gamma(k,\omega,q)=-\frac{a^2}{2\pi i}\oint_{|z|=1}\frac{dz(z^2-1)^2}{z^2[a(z^2+1)+(b+i\epsilon)z]},
\end{equation}
which according to an identity
\begin{equation}
\frac{(z^2-1)^2}{z^2[a(z^2+1)+(b+i\epsilon)z]}=\frac{1}{a}+\frac{1}{az^2}-\frac{b}{a^2z}-\frac{4a^2-b^2}{a^2[a(z^2+1)+(b+i\epsilon)z]},
\end{equation}
results in
\begin{equation}
\Gamma(k,\omega,q)=b+\frac{1}{2\pi i}\oint_{|z|=1}dz\frac{4a^2-b^2}{a(z^2+1)+(b+i\epsilon)z}.
\end{equation}

Now we are ready to get an integral representation for $\Sigma_1(\omega,q,T)$. Following (68)
\begin{equation}
\Sigma_1(\omega,q,T)=\omega-E_{magn}(q)-(G^{-1})_0-(G^{-1})_1+\Pi_1(\omega,q,T),
\end{equation}
where according to (65), (B.3) and (B.8)
\begin{eqnarray}
&&(G^{-1})_0(\omega)=\frac{\omega-\theta_0+i\epsilon}{\sigma_0}=\omega-E_{magn}+i\epsilon,\nonumber\\
&&(G^{-1})_1(\omega,q,T)=-\frac{\omega-\theta_0+i\epsilon}{\sigma_0^2}\sigma_1(T)-\frac{\theta_1(q,T)}{\sigma_0}\nonumber\\
&&=-(\omega-E_{magn}(q)+i\epsilon)\sigma_1(T)-\theta_1(q,T),
\end{eqnarray}
are the first two terms of the form factor expansion for $G^{-1}$. From (90) and (91) follows that
\begin{eqnarray}
\Sigma_1(\omega,q,T)=\Pi_1(\omega,q,T)-(\omega-E_{magn}(q)+i\epsilon)\sigma_1(T)-\theta_1(q,T),
\end{eqnarray}
or according to (84), (B.9), (86) and (89)
\begin{eqnarray}
&&\Sigma_1(\omega,q,T)=\frac{1}{\pi}\int_{-\pi}^{\pi}dk{\rm e}^{-\beta E_{magn}(k-q)}\Big(\Gamma(k,\omega,q)-b\Big)\nonumber\\
&&=\frac{1}{\pi}\int_{-\pi}^{\pi}dk{\rm e}^{-\beta E_{magn}(k-q)}\tilde\Gamma(k,\omega,q),
\end{eqnarray}
where
\begin{equation}
\tilde\Gamma(k,\omega,q)=\frac{1}{2\pi i}\oint_{|z|=1}dz\frac{4a^2-b^2}{a(z^2+1)+(b+i\epsilon)z}=\frac{4a^2-b^2}{a(z_{in}-z_{out})}.
\end{equation}
Here $z_{in}$ and $z_{out}$ are the roots of the square equation
\begin{equation}
a(z^2+1)+(b+i\epsilon)z=0,
\end{equation}
so that $z_{in}$ lies inside the unit circle
\begin{equation}
|z_{in}|<1,\qquad z_{in}z_{out}=1.
\end{equation}

According to (15), (89) and (86)
\begin{equation}
E_{magn}(k-q)=E_{magn}(q-k),\qquad\tilde\Gamma(-k,\omega,-q)=\tilde\Gamma(k,\omega,q).
\end{equation}
Hence applying the substitution $(k,q)\rightarrow(-k,-q)$ to the integrand in the right side of (93) one readily gets
\begin{equation}
\Sigma_1(\omega,-q,T)=\Sigma_1(\omega,q,T),
\end{equation}
in agreement with invariance of (1) under the inverse of the chain direction.

As it follows from (94), (95) and (86) $\tilde\Gamma(k,\omega,q)$ is real at $D(k,\omega,q)\geq0$ and pure imaginary at $D(k,\omega,q)<0$ where
\begin{equation}
D(k,\omega,q)\equiv b^2-4a^2=(\omega-\Phi_{down}(q,k))(\omega-\Phi_{up}(q,k)).
\end{equation}
and
\begin{eqnarray}
\Phi_{down}(q,k)=E_{down}(k)-E_{magn}(k-q,h)=h+|J|\Big(\cos{(k-q-k_{gap})}-2\cos{\frac{k}{2}}\Big),\nonumber\\
\Phi_{up}(q,k)=E_{up}(k)-E_{magn}(k-q,h)=h+|J|\Big(\cos{(k-q-k_{gap})}+2\cos{\frac{k}{2}}\Big).
\end{eqnarray}
In other words $\tilde\Gamma(k,\omega,q)$ is imaginary when $\omega+E_{magn}(k-q)$ lies inside the two-magnon zone and real otherwise. In the latter case
\begin{eqnarray}
&&z_{in}-z_{out}=-\frac{\sqrt{b^2-4a^2}}{a},\qquad b\leq-2|a|\Longleftrightarrow\omega\leq\Phi_{down}(q,k),\nonumber\\
&&z_{in}-z_{out}=\frac{\sqrt{b^2-4a^2}}{a},\qquad b\geq2|a|\Longleftrightarrow\omega\geq\Phi_{up}(q,k),
\end{eqnarray}
and according to (94), (99) and (101)
\begin{eqnarray}
&&\tilde\Gamma(k,\omega,q)=\sqrt{D(k,\omega,q)},\qquad\omega\leq\Phi_{down}(q,k),\nonumber\\
&&\tilde\Gamma(k,\omega,q)=-\sqrt{D(k,\omega,q)},\qquad\omega\geq\Phi_{up}(q,k).
\end{eqnarray}
At $\Phi_{down}(q,k)<\omega<\Phi_{up}(q,k)$ when the function $\tilde\Gamma(k,\omega,q)$ is purely imaginary
one may back from $z$ to $\kappa$. Then according to (94), (25) and (99)
\begin{eqnarray}
&&\tilde\Gamma(k,\omega,q)=-i\frac{D(k,\omega,q)}{\pi}{\rm Im}\int_0^{\pi}\frac{d\kappa}{2a\cos{\kappa}+b+i\epsilon}=i\frac{D(k,\omega,q)}{\sqrt{4a^2-b^2}}\nonumber\\
&&=-i\sqrt{|D(k,\omega,q)|},\qquad\Phi_{down}(q,k)<\omega<\Phi_{up}(q,k).
\end{eqnarray}
Now gathering together (93), (102) and (103) one readily gets the representation
\begin{eqnarray}
&&\Sigma_1(\omega,q,T)=\frac{1}{\pi}\int_{-\pi}^{\pi}dk{\rm e}^{-\beta E_{magn}(k-q)}\sqrt{|D(k,\omega,q)|}\Big[\Theta(\Phi_{down}(q,k)-\omega)\nonumber\\
&&-\Theta(\omega-\Phi_{up}(q,k))-i\Theta(\Phi_{up}(q,k)-\omega)\Theta(\omega-\Phi_{down}(q,k))\Big],
\end{eqnarray}
where $\Theta(x)$ is the Heaviside function. The compact formula (104) is the most general result of the paper. However for three separate diapasons $\omega<E_{magn}(q)$, $\omega=E_{magn}(q)$
and $\omega>E_{magn}(q)$ the number of $\Theta$-functions may be reduced.

First of all let us notice that according to (15) (100 )and (39)
\begin{eqnarray}
&&E_{magn}(q)-\Phi_{down}(q,k)=2|J|\cos{\frac{k}{2}}\Big[1-\cos{\Big(\frac{k}{2}-q-k_{gap}\Big)}\Big]\geq0,\nonumber\\
&&\Phi_{up}(q,k)-E_{magn}(q)=2|J|\cos{\frac{k}{2}}\Big[1+\cos{\Big(\frac{k}{2}-q-k_{gap}\Big)}\Big]\geq0.
\end{eqnarray}
Hence for $\omega=E_{magn}(q)$ (104) reduces to
\begin{equation}
\Sigma_1(E_{magn}(q),q,T)=-\frac{i}{\pi}\int_{-\pi}^{\pi}dk{\rm e}^{-\beta E_{magn}(k)}\sqrt{|D(k+q,E_{magn}(q),q)|}.
\end{equation}
At the same time according to (99) and (105)
\begin{equation}
\sqrt{|D(k+q,E_{magn}(q),q)|}=2|J|\Big|\cos{\frac{k+q}{2}}\sin{\frac{k-q}{2}}\Big|=|J(\sin{k}-\sin{q})|.
\end{equation}
Hence (106) further reduces to
\begin{equation}
\Sigma_1(E_{magn}(q),q,T)=-\frac{i|J|}{\pi}\int_{-\pi}^{\pi}dk{\rm e}^{-\beta E_{magn}(k)}|\sin{k}-\sin{q}|.
\end{equation}

Turning to the cases $\omega<E_{magn}(q)$ and $\omega>E_{magn}(q)$, let us first prove that
\begin{equation}
{\rm e}^{-\beta E_{gap}}\ll1,\quad \omega\not\in[\omega_{min}(q),\omega_{max}(q)]\Longrightarrow S(\omega,q,T)=0,
\end{equation}
where $\omega_{min}(q)$ and $\omega_{max}(q)$ are correspondingly the minimal value of $\Phi_{down}(q,k)$ and the maximal value of $\Phi_{up}(q,k)$ for $k\in[-\pi,\pi]$.
Namely as it is shown in the Appendix B (for $q\in[-\pi,\pi]$)
\begin{equation}
\omega_{min}(q)=h-3|J|\cos{\frac{|q|+k_{gap}-\pi}{3}},\qquad\omega_{max}(q)=h+3|J|\cos{\frac{|q|-k_{gap}}{3}}.
\end{equation}
Representing (15) in two equivalent forms
\begin{equation}
E_{magn}(q)=h+|J|\cos{(|q|+k_{gap}-\pi)}=h-|J|\cos{(|q|-k_{gap})},
\end{equation}
and using the well known formula $\cos{3x}=4\cos^3{x}-3\cos{x}$ one readily gets
\begin{eqnarray}
&&E_{magn}(q)-\omega_{min}(q)=4|J|\cos^3{\frac{|q|+k_{gap}-\pi}{3}}\geq4|J|\cos^3{\frac{\pi}{3}}=\frac{|J|}{2},\nonumber\\
&&\omega_{max}(q)-E_{magn}(q)=4|J|\cos^3{\frac{|q|-k_{gap}}{3}}\geq4|J|\cos^3{\frac{\pi}{3}}=\frac{|J|}{2}.
\end{eqnarray}
Hence
\begin{equation}
\omega\not\in[\omega_{min}(q),\omega_{max}(q)]\Longrightarrow|\omega-E_{magn}(q)|\geq\frac{|J|}{2}.
\end{equation}
At the same time according to (104) in this case ($\omega\not\in[\omega_{min}(q),\omega_{max}(q)]$) one has ${\rm Im}\Sigma_1(\omega,q,T)=0$ so that
\begin{equation}
S(\omega,q,T)=\delta(\omega-E_{magn}(q)-\Sigma_1(\omega,q,T)),\qquad\Sigma_1(\omega,q,T)={\rm Re}\Sigma_1(\omega,q,T).
\end{equation}
But from (113) and the relation $\Sigma_1(\omega,q,T)=O(|J|{\rm e}^{-\beta E_{magn}})$ follows that the equation
\begin{equation}
\omega-E_{magn}(q)-\Sigma_1(\omega,q,T)=0,
\end{equation}
has no solutions at small $T$ and $\omega\not\in[\omega_{min}(q),\omega_{max}(q)]$. This proves (109).

Using now (105) and (109) we may reduce (104) considering it separately in the two diapasons $\omega_{min}(q)\leq\omega<E_{magn}(q)$ and $E_{magn}(q)<\omega\leq\omega_{max}(q)$. Namely
\begin{eqnarray}
&&\Sigma_1(\omega,q,T)=\frac{1}{\pi}\int_{-\pi}^{\pi}dk{\rm e}^{-\beta E_{magn}(k-q)}\sqrt{|D(k,\omega,q)|}\Big[\Theta(\Phi_{down}(q,k)-\omega)\nonumber\\
&&-i\Theta(\omega-\Phi_{down}(q,k))\Big],\qquad\omega_{min}(q)\leq\omega<E_{magn}(q),\\
&&\Sigma_1(\omega,q,T)=-\frac{1}{\pi}\int_{-\pi}^{\pi}dk{\rm e}^{-\beta E_{magn}(k-q)}\sqrt{|D(k,\omega,q)|}\Big[\Theta(\omega-\Phi_{up}(q,k))\nonumber\\
&&+i\Theta(\Phi_{up}(q,k)-\omega)\Big],\qquad E_{magn}(q)<\omega\leq\omega_{max}(q).
\end{eqnarray}
In the next three sections for three special cases $q=0,\pi/2,\pi$ we shall further simplify these expressions and reduce all the $\Theta$-functions.

\section{Low temperature asymptotic of TDSF at $q=k_{gap}$}

According to (100) the functions
\begin{eqnarray}
&&\Phi_{down}(k_{gap},k)=\Phi_{down}(k_{gap},-k)=E_{gap}+E_{width}\Big(\cos^2{\frac{k}{2}}-\cos{\frac{k}{2}}\Big),\nonumber\\
&&\Phi_{up}(k_{gap},k)=\Phi_{down}(k_{gap},-k)=E_{gap}+E_{width}\Big(\cos^2{\frac{k}{2}}+\cos{\frac{k}{2}}\Big),
\end{eqnarray}
are even. Here
\begin{equation}
E_{width}=2|J|,
\end{equation}
is the magnon band width.

The function $\Phi_{down}(k_{gap},k)$ has two equal symmetric minima at $k=\pm\pi/3$ while $\Phi_{up}(k_{gap},k)$ has a single maximum at $k=0$.
A substitution of (118)  into (99) yields
\begin{equation}
D(k,\omega,k_{gap})=E_{width}^2\Big[\Big(\frac{\omega-E_{gap}}{E_{width}}-\cos^2{\frac{k}{2}}\Big)^2-\cos^2{\frac{k}{2}}\Big].
\end{equation}
From (15), (2) and (119) follow that
\begin{equation}
E_{magn}(k-k_{gap})=h-|J|\cos{k}=E_{gap}+E_{width}\sin^2{\frac{k}{2}}.
\end{equation}
According to (15) and (110)
\begin{equation}
E_{magn}(k_{gap})=E_{gap}=h-|J|,\quad\omega_{min}(k_{gap})=h-\frac{3|J|}{2},\quad\omega_{max}(k_{gap})=h+3|J|.
\end{equation}

Using (120)-(122) one reduce (116) and (117) to
\begin{eqnarray}
&&{\rm Re}\Sigma_1(\omega,k_{gap},T)=\frac{2E_{width}{\rm e}^{-\beta E_{gap}}}{\pi}\Big(\int_0^{k_{down}^+}dk+\int_{k_{down}^-}^{\pi}dk\Big){\rm e}^{-\beta E_{width}\sin^2{k/2}}\nonumber\\
&&\cdot\sqrt{\Big(\frac{\omega-E_{gap}}{E_{width}}-\cos^2{\frac{k}{2}}\Big)^2-\cos^2{\frac{k}{2}}},\nonumber\\
&&{\rm Im}\Sigma_1(\omega,k_{gap},T)=-\frac{2E_{width}{\rm e}^{-\beta E_{gap}}}{\pi}\int_{k_{down}^+}^{k_{down}^-}dk{\rm e}^{-\beta E_{width}\sin^2{k/2}}\nonumber\\
&&\cdot\sqrt{\cos^2{\frac{k}{2}}-\Big(\frac{\omega-E_{gap}}{E_{width}}-\cos^2{\frac{k}{2}}\Big)^2},\qquad\omega_{min}(k_{gap})\leq\omega<E_{gap},
\end{eqnarray}
and
\begin{eqnarray}
&&{\rm Re}\Sigma_1(\omega,k_{gap},T)=-\frac{2E_{width}{\rm e}^{-\beta E_{gap}}}{\pi}\int_{k_{up}}^{\pi}dk{\rm e}^{-\beta E_{width}\sin^2{k/2}}\nonumber\\
&&\cdot\sqrt{\Big(\frac{\omega-E_{gap}}{E_{width}}-\cos^2{\frac{k}{2}}\Big)^2-\cos^2{\frac{k}{2}}},\nonumber\\
&&{\rm Im}\Sigma_1(\omega,k_{gap},T)=-\frac{2E_{width}{\rm e}^{-\beta E_{gap}}}{\pi}\int_0^{k_{up}}dk{\rm e}^{-\beta E_{width}\sin^2{k/2}}\nonumber\\
&&\cdot\sqrt{\cos^2{\frac{k}{2}}-\Big(\frac{\omega-E_{gap}}{E_{width}}-\cos^2{\frac{k}{2}}\Big)^2},\qquad E_{gap}<\omega\leq\omega_{max}(k_{gap}).
\end{eqnarray}
According to the evenness of integrands the integrals in (123) and (124) are taken only over positive $k$.

The two boundaries $0\leq k_{down}^+\leq k_{down}^-\leq\pi$ in (123) are the two solutions of the equation
\begin{equation}
\omega-\Phi_{down}(k_{gap},k_{down})=0,
\end{equation}
which under a substitution of (118) takes the form
\begin{equation}
x_{down}^2-x_{down}-\lambda_+=0,\qquad x_{down}\equiv\cos{\frac{k_{down}}{2}},\qquad\lambda_+=\frac{\omega-E_{gap}}{E_{width}}.
\end{equation}
Solving (126) one readily gets
\begin{equation}
k_{down}^{\pm}=2\arccos{\Big(\frac{1\pm\sqrt{1+4\lambda_+}}{2}\Big)},\qquad-\frac{1}{4}\leq\lambda_+<0\Leftrightarrow\omega_{min}(k_{gap})\leq\omega<E_{gap}.
\end{equation}

The boundary $k_{up}\geq0$ in (124) is the positive solution of the equation
\begin{equation}
\omega-\Phi_{up}(k_{gap},k_{up})=0,
\end{equation}
A substitution of (118) reduces (128) to
\begin{equation}
x_{up}^2+x_{up}-\lambda_+=0,\qquad x_{up}\equiv\cos{\frac{k_{up}}{2}},\qquad\lambda_+=\frac{\omega-E_{gap}}{E_{width}},
\end{equation}
and yields
\begin{equation}
k_{up}=2\arccos{\Big(\frac{\sqrt{1+4\lambda_+}-1}{2}\Big)},\qquad0<\lambda_+\leq2\Leftrightarrow E_{gap}<\omega\leq\omega_{max}(k_{gap}).
\end{equation}

Evaluation of $k^{\pm}_{down}$ and $k_{up}$ at $h=2.5$ and $J=0.5$ on the base of (125) and (128) is graphically illustrated on Fig. 1. The corresponding line shapes for various $\beta E_{gap}$ are presented on Fig. 2.

\section{Low temperature asymptotic of TDSF at $q=\pi-k_{gap}$}

This case is dual to the one considered in the previous section. According to (100) and (15) the functions
\begin{eqnarray}
&&\Phi_{down}(\pi-k_{gap},k)=E_{magn}(\pi-k_{gap})-E_{width}\Big(\cos^2{\frac{k}{2}}+\cos{\frac{k}{2}}\Big),\nonumber\\
&&\Phi_{up}(\pi-k_{gap},k)=E_{magn}(\pi-k_{gap})-E_{width}\Big(\cos^2{\frac{k}{2}}-\cos{\frac{k}{2}}\Big),
\end{eqnarray}
also are even. $\Phi_{down}(k_{gap},k)$ has a single minimum at $k=0$, while $\Phi_{up}(k_{gap},k)$ has two symmetric maxima at $k=\pm\pi/3$.
A substitution of (131)  into (99) yields
\begin{equation}
D(k,\omega,\pi-k_{gap})=E_{width}^2\Big[\Big(\frac{\omega-E_{magn}(\pi-k_{gap})}{E_{width}}+\cos^2{\frac{k}{2}}\Big)^2-\cos^2{\frac{k}{2}}\Big].
\end{equation}
From (15), (2) and (119) follows
\begin{equation}
E_{magn}(k-k_{gap}+\pi)=h+|J|\cos{k}=E_{gap}+E_{width}\cos^2{\frac{k}{2}}.
\end{equation}
According to (15) and (110)
\begin{equation}
E_{magn}(\pi-k_{gap})=h+|J|,\quad\omega_{min}(\pi-k_{gap})=h-3|J|,\quad\omega_{max}(\pi-k_{gap})=h+\frac{3|J|}{2}.
\end{equation}

Using (132), (133) one reduces (116) and (117) to
\begin{eqnarray}
&&{\rm Re}\Sigma_1(\omega,\pi-k_{gap},T)=\frac{2E_{width}{\rm e}^{-\beta E_{gap}}}{\pi}\int_{k_{down}}^{\pi}dk{\rm e}^{-\beta E_{width}\cos^2{k/2}}\nonumber\\
&&\cdot\sqrt{\Big(\frac{\omega-E_{magn}(\pi-k_{gap})}{E_{width}}+\cos^2{\frac{k}{2}}\Big)^2-\cos^2{\frac{k}{2}}},\nonumber\\
&&{\rm Im}\Sigma_1(\omega,\pi-k_{gap},T)=-\frac{2E_{width}{\rm e}^{-\beta E_{gap}}}{\pi}\int_0^{k_{down}}dk{\rm e}^{-\beta E_{width}\cos^2{k/2}}\nonumber\\
&&\cdot\sqrt{\cos^2{\frac{k}{2}}-\Big(\frac{\omega-E_{magn}(\pi-k_{gap})}{E_{width}}+\cos^2{\frac{k}{2}}\Big)^2},\nonumber\\
&&\omega_{min}(\pi-k_{gap})<\omega<E_{magn}(\pi-k_{gap}),
\end{eqnarray}
and
\begin{eqnarray}
&&{\rm Re}\Sigma_1(\omega,\pi-k_{gap},T)=-\frac{2E_{width}{\rm e}^{-\beta E_{gap}}}{\pi}\Big(\int_0^{k_{up}^+}dk+\int_{k_{up}^-}^{\pi}dk\Big){\rm e}^{-\beta E_{width}\cos^2{k/2}}\nonumber\\
&&\cdot\sqrt{\Big(\frac{\omega-E_{magn}(\pi-k_{gap})}{E_{width}}+\cos^2{\frac{k}{2}}\Big)^2-\cos^2{\frac{k}{2}}},\nonumber\\
&&{\rm Im}\Sigma_1(\omega,\pi-k_{gap},T)=-\frac{2E_{width}{\rm e}^{-\beta E_{gap}}}{\pi}\int_{k_{up}^+}^{k_{up}^-}dk{\rm e}^{-\beta E_{width}\cos^2{k/2}}\nonumber\\
&&\cdot\sqrt{\cos^2{\frac{k}{2}}-\Big(\frac{\omega-E_{magn}(\pi-k_{gap})}{E_{width}}+\cos^2{\frac{k}{2}}\Big)^2},\nonumber\\
&&E_{magn}(\pi-k_{gap})<\omega<\omega_{max}(\pi-k_{gap}).
\end{eqnarray}
Due to the evenness of integrands the integrals in (135) and (136) are taken only over positive $k$.

The boundary $k_{down}\geq0$ in (135) is the positive solution of the equation
\begin{equation}
\omega-\Phi_{down}(k_{gap},k_{down})=0.
\end{equation}
Rewriting (137) in the equivalent form with the use of (131)
\begin{equation}
x_{down}^2+x_{down}+\lambda_-=0,\qquad x_{down}\equiv\cos{\frac{k_{down}}{2}},\qquad\lambda_-=\frac{\omega-E_{magn}(\pi-k_{gap})}{E_{width}},
\end{equation}
one readily gets
\begin{eqnarray}
&&k_{down}=2\arccos{\Big(\frac{\sqrt{1-4\lambda_-}-1}{2}\Big)},\nonumber\\
&&-2\leq\lambda_-<0\Leftrightarrow\omega_{min}(\pi-k_{gap})\leq\omega<E_{magn}(\pi-k_{gap}).
\end{eqnarray}

The two boundaries $0\leq k_{up}^+\leq k_{up}^-\leq\pi$ in (136) are the two solutions of the equation
\begin{equation}
\omega-\Phi_{up}(k_{gap},k_{up})=0.
\end{equation}
A substitution of (131) reduces (140) to
\begin{equation}
x_{up}^2-x_{up}+\lambda_-=0,\qquad x_{up}\equiv\cos{\frac{k_{up}}{2}},\qquad\lambda_-=\frac{\omega-E_{magn}(\pi-k_{gap})}{E_{width}},
\end{equation}
and yields
\begin{eqnarray}
&&k_{up}^{\pm}=2\arccos{\Big(\frac{1\pm\sqrt{1-4\lambda_-}}{2}\Big)},\nonumber\\
&&0<\lambda_-\leq\frac{1}{4}\Leftrightarrow E_{magn}(\pi-k_{gap})<\omega\leq\omega_{max}(\pi-k_{gap}).
\end{eqnarray}

Evaluation of $k_{down}$ and $k^{\pm}_{up}$ at $h=2.5$ and $J=0.5$ on the base of (137) and (140) is graphically illustrated on Fig. 3. The corresponding line shapes for various $\beta E_{gap}$ are presented on Fig. 4.

\section{Low temperature asymptotic of TDSF at $|q|=\pi/2$}

Since the model (1) is invariant under the spatial inversion there should be
\begin{equation}
S(\omega,-q,T)=S(\omega,q,T).
\end{equation}
Hence evaluating $S(\omega,q,T)$ at
\begin{equation}
q=\frac{\pi}{2}-k_{gap},
\end{equation}
we additionally to the pair of cases $k_{gap}=0$, $q=\pi/2$ and $k_{gap}=\pi$, $q=-\pi/2$ study the dual one which is $k_{gap}=0$, $q=-\pi/2$ and $k_{gap}=\pi$, $q=\pi/2$. For all of them $|q|=\pi/2$.

A substitution of (144) into (100) yields
\begin{eqnarray}
&&\Phi_{down}(\pi/2-k_{gap},k)=h+|J|\Big(\sin{k}-2\cos{\frac{k}{2}}\Big),\nonumber\\
&&\Phi_{up}(\pi/2-k_{gap},k)=h+|J|\Big(\sin{k}+2\cos{\frac{k}{2}}\Big).
\end{eqnarray}
It may be readily proved that for $-\pi<k\leq\pi$ the function $\Phi_{down}(\pi/2-k_{gap},k)$ has a single minimum at $k=-\pi/3$, while the function $\Phi_{up}(\pi/2-k_{gap},k)$ has a single maximum
at $k=\pi/3$. Hence for $\omega$ in the intervals
\begin{eqnarray}
h-\frac{3\sqrt{3}|J|}{2}=h-3|J|\cos{\frac{\pi}{6}}=\omega_{min}\Big(\frac{\pi}{2}-k_{gap}\Big)\leq\omega\leq E_{magn}\Big(\frac{\pi}{2}-k_{gap}\Big)=h,\\
h=E_{magn}\Big(\frac{\pi}{2}-k_{gap}\Big)\leq\omega\leq\omega_{max}\Big(\frac{\pi}{2}-k_{gap}\Big)=h+3|J|\cos{\frac{\pi}{6}}=h+\frac{3\sqrt{3}|J|}{2},
\end{eqnarray}
both the equations
\begin{eqnarray}
&&\omega-\Phi_{down}(\pi/2-k_{gap},k_{down})=0,\qquad h-\frac{3\sqrt{3}|J|}{2}\leq\omega\leq h,\\
&&\omega-\Phi_{up}(\pi/2-k_{gap},k_{up})=0\qquad h\leq\omega\leq h+\frac{3\sqrt{3}|J|}{2},
\end{eqnarray}
have exactly two solutions $-\pi\leq k_{down}^{(1)}<k_{down}^{(2)}\leq\pi$ and $-\pi\leq k_{up}^{(1)}<k_{up}^{(2)}\leq\pi$.

Taking
\begin{equation}
\lambda_0\equiv\frac{\omega-E_{magn}(\pi/2-k_{gap})}{E_{width}}=\frac{\omega-h}{2|J|},
\end{equation}
we rewrite (148), (149) in the forms
\begin{eqnarray}
&&\lambda_0=\cos{\frac{k_{down}}{2}}\Big(\sin{\frac{k_{down}}{2}}-1\Big),\qquad-\frac{3\sqrt{3}}{4}\leq\lambda_0<0,\\
&&\lambda_0=\cos{\frac{k_{up}}{2}}\Big(\sin{\frac{k_{up}}{2}}+1\Big),\qquad0\leq\lambda_0\leq\frac{3\sqrt{3}}{4},
\end{eqnarray}
Under a substitution
\begin{equation}
k_{down}\longrightarrow-k_{up},\qquad\lambda_0\longrightarrow-\lambda_0,
\end{equation}
(151) turns into (152). It is convenient to transform (151) and (152) into the following quartic equations
\begin{eqnarray}
&&(x_{down}^2-1)(x_{down}-1)^2+\lambda_0^2=0,\qquad-\frac{3\sqrt{3}}{4}\leq\lambda_0\leq0,\\
&&(x_{up}^2-1)(x_{up}+1)^2+\lambda_0^2=0,\qquad0\leq\lambda_0\leq\frac{3\sqrt{3}}{4},
\end{eqnarray}
where $x_{down}\equiv\sin{k_{down}/2}$ and $x_{up}\equiv\sin{k_{up}/2}$. The symmetry (153) reduces now to
\begin{equation}
x_{down}\longrightarrow-x_{up},
\end{equation}
and turns (154) into (155).

As it is shown in the Appendix D (154) has only two real solutions
\begin{equation}
x_{\pm}=\frac{1}{2}\left(1-\sqrt{t+1}\pm\sqrt{2-t+\frac{2}{\sqrt{t+1}}}\right),
\end{equation}
where
\begin{equation}
t=\sqrt[3]{2\lambda_0^2}\left(\sqrt[3]{1+\sqrt{1-\frac{16\lambda_0^2}{27}}}+\sqrt[3]{1-\sqrt{1-\frac{16\lambda_0^2}{27}}}\right).
\end{equation}
According to the symmetry (156)
\begin{equation}
k^{(1)}_{down}=k_-,\qquad k^{(2)}_{down}=k_+,\qquad k^{(1)}_{up}=-k_+,\qquad k^{(2)}_{up}=-k_-,
\end{equation}
where
\begin{equation}
k_{\pm}=2\arcsin{x_{\pm}}.
\end{equation}

According to (15), (2), (99) and (145) one readily has
\begin{eqnarray}
&&E_{magn}\Big(k+k_{gap}-\frac{\pi}{2}\Big)=h-|J|\sin{k}=E_{gap}+\frac{E_{width}(1-\sin{k})}{2},\\
&&D\Big(k,\omega,\frac{\pi}{2}-k_{gap}\Big)=E_{width}^2\Big[\Big(\frac{\omega-h}{E_{width}}-\frac{1}{2}\sin{k}\Big)^2-\cos^2{\frac{k}{2}}\Big].
\end{eqnarray}
Hence (116) and (117) reduce to
\begin{eqnarray}
&&{\rm Re}\Sigma_1(\omega,\pm\pi/2,T)=\frac{E_{width}{\rm e}^{-\beta E_{gap}}}{\pi}\Big(\int_{-\pi}^{k_-}dk+\int_{k_+}^{\pi}dk\Big){\rm e}^{-\beta(E_{width}/2)(1-\sin{k})}\nonumber\\
&&\cdot\sqrt{\Big(\frac{\omega-h}{E_{width}}-\frac{1}{2}\sin{k}\Big)^2-\cos^2{\frac{k}{2}}},\nonumber\\
&&{\rm Im}\Sigma_1(\omega,k_{gap},T)=-\frac{E_{width}{\rm e}^{-\beta E_{gap}}}{\pi}\int_{k_-}^{k_+}dk{\rm e}^{-\beta(E_{width}/2)(1-\sin{k})}\nonumber\\
&&\cdot\sqrt{\cos^2{\frac{k}{2}}-\Big(\frac{\omega-h}{E_{width}}-\frac{1}{2}\sin{k}\Big)^2},\qquad\omega_{min}(\pi/2-k_{gap})\leq\omega<h,
\end{eqnarray}
and
\begin{eqnarray}
&&{\rm Re}\Sigma_1(\omega,\pm\pi/2,T)=-\frac{E_{width}{\rm e}^{-\beta E_{gap}}}{\pi}\Big(\int_{-\pi}^{-k_+}dk+\int_{-k_-}^{\pi}dk\Big){\rm e}^{-\beta(E_{width}/2)(1-\sin{k})}\nonumber\\
&&\cdot\sqrt{\Big(\frac{\omega-h}{E_{width}}-\frac{1}{2}\sin{k}\Big)^2-\cos^2{\frac{k}{2}}},\nonumber\\
&&{\rm Im}\Sigma_1(\omega,k_{gap},T)=-\frac{E_{width}{\rm e}^{-\beta E_{gap}}}{\pi}\int_{-k_+}^{-k_-}dk{\rm e}^{-\beta(E_{width}/2)(1-\sin{k})}\nonumber\\
&&\cdot\sqrt{\cos^2{\frac{k}{2}}-\Big(\frac{\omega-h}{E_{width}}-\frac{1}{2}\sin{k}\Big)^2},\qquad h<\omega\leq\omega_{max}(\pi/2-k_{gap}).
\end{eqnarray}

Evaluation of $k_{down}^{(1,2)}$ and $k^{(1,2)}_{up}$ at $h=2.5$ and $J=0.5$ on the base of (148) and (149) is graphically illustrated on Fig. 5. The corresponding line shapes for various $\beta E_{gap}$ are presented on Fig. 6.

\section{Summary and discussion}

In the present paper we have derived the integral representation (104) (or in a more transparent form (116) and (117)) for the low-temperature asymptotic of the magnon self energy in the model (1).
Its substitution into the Dyson representation (26) results in the low-temperature asymptotic for the dynamical magnetic susceptibility which in its turn according to (18) gives
the corresponding asymptotic for $S(\omega,q,T)$ at
$\omega\neq0$. At the special values $|q|=0,\pi/2,\pi$ the expressions for
(116) and (117) were further simplified and the corresponding line shapes were presented.

The progress originates form the use of two different approaches. The former one suggested by N. M. Plakida and Yu. A. Tserkovnikov \cite{19,20,21} allows to rigorously
obtain the Dyson equation. On this base the latter one \cite{13,14,15} allows to obtain low-temperature asymptotics for the self energy, dynamical magnetic susceptibility and TDSF.
Since the suggested approach is an "alloy" of the two already pointed ones it has not only similarities but also differences with both of them. Namely.

\begin{itemize}
\item In \cite{19,20,21} the zero order Green function was suggested in the temperature dependent Roth variational form \cite{23} (see (65) and (67)). In the present paper we follow this method only on the first
stage and then turn from the Roth Green function (65) to the zero temperature one (24). Correspondingly obtaining on the first stage the Dyson representation in the form (64) suggested in \cite{19,20,21}
we then transform it into the form (26) used in \cite{13,14,15} as the starting point for the form factor expansion.
\item The authors of \cite{13,14,15} used the temperature (Matsubara) Green function for which the Dyson equation may be proved only perturbatively. Moreover an exact form of the Dyson equation
for spin models is still under discussion \cite{17,18}. In \cite{13,14,15} the analog of equation (26) for the temperature Green function was {\it postulated}. At the same time in the present paper
we use the real two time retarded Green function (19)-(21) for which the Dyson equation may be rigorously proved just within the approach \cite{19,20,21}.
\end{itemize}

Some important aspects were not elucidated in the paper. First of all we did not considered the zero-frequency anomaly term $C\delta(\omega)$ which should de added to (18) if we remove the condition
$\omega\neq0$ \cite{24}. The constant $C$ has a clear physical meaning and corresponds both to the difference between isothermal and isolated static susceptibilities and to ergodic
properties of the system.
We did not compared our results with the corresponding ones related to the space-time Green functions $g(t,n,T)$ (5), (9) \cite{4,7}.
To our opinion before doing this it will be useful to obtain the low-temperature expansion for $g(t,n,T)$ on the base of the approach developed in \cite{25} (which in fact is similar to the one used in \cite{19,20,21}).
The author hopes to study all these problems in future.

\appendix
\renewcommand{\theequation}{\thesection.\arabic{equation}}

\section{Some formulas related to Green functions}
\setcounter{equation}{0}

According to (8) and an obvious relation
\begin{equation}
\langle\nu|{\bf S}^+(q)|\mu\rangle^*=\langle\mu|{\bf S}^-(-q)|\nu\rangle,
\end{equation}
one has
\begin{eqnarray}
&&\frac{1}{N}\sum_q{\rm e}^{inq}|\langle\nu|{\bf S}^+(q)|\mu\rangle|^2=\frac{1}{N}\sum_q{\rm e}^{inq}\langle\nu|{\bf S}^+(q)|\mu\rangle\langle\mu|{\bf S}^-(-q)|\nu\rangle\nonumber\\
&&=\frac{1}{N^2}\sum_{q,n_1,n_2}{\rm e}^{i(n-n_2+n_1)q}\langle\nu|{\bf S}_{n_2}^+|\mu\rangle\langle\mu|{\bf S}^-_{n_1}|\nu\rangle=
\langle\nu|{\bf S}_n^+|\mu\rangle\langle\mu|{\bf S}^-_0|\nu\rangle.
\end{eqnarray}
Hence (9) may be proved as follows
\begin{eqnarray}
&&\frac{1}{2\pi}\int_{-\infty}^{\infty}d\omega\int_{-\pi}^{\pi}dq{\rm e}^{i(qn-\omega t)}S(\omega,q,T)=\int_{-\infty}^{\infty}d\omega{\rm e}^{-i\omega t}\lim_{N\rightarrow\infty}\frac{1}{N}
\sum_q{\rm e}^{iqn}S(\omega,q,T)\nonumber\\
&&=\lim_{N\rightarrow\infty}\frac{1}{Z(\beta,N)}\sum_{\mu,\nu}{\rm e}^{-\beta E_{\nu}}{\rm e}^{i(E_{\nu}-E_{\mu})t}\langle\nu|{\bf S}_n^+|\mu\rangle\langle\mu|{\bf S}^-_0|\nu\rangle\nonumber\\
&&=\lim_{N\rightarrow\infty}\frac{1}{Z(\beta,N)}\sum_{\mu,\nu}{\rm e}^{-\beta E_{\nu}}\langle\nu|{\bf S}_n^+(t)|\mu\rangle\langle\mu|{\bf S}^-_0|\nu\rangle=g(t,n,T).
\end{eqnarray}

Taking the spectral representation for the commutator
\begin{equation}
\langle[A(t),B]\rangle=\sum_{\mu,\nu}\frac{{\rm e}^{i(E_{\nu}-E_{\mu})t}}{Z(T,N)}\Big({\rm e}^{-\beta E_{\nu}}\langle\nu|A|\mu\rangle\langle\mu|B|\nu\rangle-{\rm e}^{-\beta E_{\mu}}\langle\mu|B|\nu\rangle\langle\nu|A|\mu\rangle\Big),
\end{equation}
which directly follows from the formula $A(t)={\rm e}^{i\hat Ht}A{\rm e}^{-i\hat Ht}$ one readily gets the spectral decomposition
\begin{equation}
\langle\langle A,B\rangle\rangle_{\omega}=\frac{1}{Z(T,N)}\sum_{\mu,\nu}\frac{{\rm e}^{-\beta E_{\nu}}-{\rm e}^{-\beta E_{\mu}}}{\omega+E_{\nu}-E_{\mu}+i\epsilon}
\langle\nu|A|\mu\rangle\langle\mu|B|\nu\rangle.
\end{equation}
Formula (23) follows now from (A.5) and (A.1).

\section{Evaluation of $\sigma_1(T)$ and $\theta_1(q,T)$}

\setcounter{equation}{0}

Since
\begin{equation}
{\bf S}^z_n|\emptyset\rangle=\frac{1}{2}|\emptyset\rangle,\qquad\sum_{n=1}^N\Big(\frac{1}{2}-{\bf S}^z_n\Big)|k\rangle=|k\rangle
\Longrightarrow\sum_{n=1}^N\langle k|{\bf S}_n^z|k\rangle=\frac{N}{2}-1,
\end{equation}
one has
\begin{eqnarray}
&&\sigma(T,N)=\frac{2\Big(\sum_{n=1}^N\langle\emptyset|{\bf S}_n^z|\emptyset\rangle+\sum_k{\rm e}^{-\beta E_{magn}(k)}\sum_{n=1}^N\langle k|{\bf S}_n^z|k\rangle\Big)+o\Big({\rm e}^{-\beta E_{gap}}\Big)}
{N\Big(1+\sum_k{\rm e}^{-\beta E_{magn}(k)}\Big)+o\Big({\rm e}^{-\beta E_{gap}}\Big)},
\nonumber\\
&&=\frac{1+\Big(1-\frac{2}{N}\Big)\sum_k{\rm e}^{-\beta E_{magn}(k)}+o\Big({\rm e}^{-\beta E_{gap}}\Big)}{1+\sum_k{\rm e}^{-\beta E_{magn}(k)}+o\Big({\rm e}^{-\beta E_{gap}}\Big)}\nonumber\\
&&=1-\frac{2}{N}\sum_k{\rm e}^{-\beta E_{magn}(k)}+o\Big({\rm e}^{-\beta E_{gap}}\Big).
\end{eqnarray}
From (B.2) follows that
\begin{equation}
\sigma_0=1,\qquad\sigma_1(T)=-\lim_{N\rightarrow\infty}\frac{2}{N}\sum_k{\rm e}^{-\beta E_{magn}(k)}=-\frac{1}{\pi}\int_0^{2\pi}dk{\rm e}^{-\beta E_{magn}(k)}.
\end{equation}

Using the explicit form of one magnon state (13) one may readily prove that
\begin{equation}
\sum_{n=1}^N\langle k|{\bf S}_n^z{\bf S}_{n+1}^z|k\rangle=\frac{N-2}{4}-\frac{2}{4}=\frac{N}{4}-1.
\end{equation}
So according to (53), (B.1) and (B.4)
\begin{equation}
\langle\emptyset|V(q)|\emptyset\rangle=-J\cos{q},\qquad
\langle k|V(q)|k\rangle=J\Big(-\cos{q}+\frac{2\cos{k}+4\cos{q}}{N}\Big),
\end{equation}
and correspondingly
\begin{eqnarray}
&&v(q,T,N)=\frac{\langle\emptyset|V(q)|\emptyset\rangle+\sum_k{\rm e}^{-\beta E_{magn}(k)}\langle k|V(q)|k\rangle+o\Big({\rm e}^{-\beta E_{gap}}\Big)}
{1+\sum_k{\rm e}^{-\beta E_{magn}(k)}+o\Big({\rm e}^{-\beta E_{gap}}\Big)}\nonumber\\
&&=\langle\emptyset|V(q)|\emptyset\rangle+\sum_k{\rm e}^{-\beta E_{magn}(k)}\Big(\langle k|V(q)|k\rangle-\langle\emptyset|V(q)|\emptyset\rangle\Big)+o\Big({\rm e}^{-\beta E_{gap}}\Big)\nonumber\\
&&=J\Big(-\cos{q}+\frac{2}{N}\sum_k{\rm e}^{-\beta E_{magn}(k)}(\cos{k}+2\cos{q})\Big)+o\Big({\rm e}^{-\beta E_{gap}}\Big).
\end{eqnarray}
Using the standard substitution $\sum_k\rightarrow N/(2\pi)\int_0^{2\pi}dk$ one readily gets from (B.6)
\begin{equation}
v_0(q,T)=-J\cos{q},\qquad v_1(q,T)=\frac{J}{\pi}\int_0^{2\pi}dk{\rm e}^{-\beta E_{magn}(k)}(\cos{k}+2\cos{q}).
\end{equation}
Now according to (54)
\begin{eqnarray}
&&\theta_0(q,T)=h+\frac{v_0(q,T)}{\sigma_0(T)}=E_{magn}(q),\nonumber\\
&&\theta_1(q,T)=\frac{v_1(q,T)}{\sigma_0(T)}-\frac{v_0(q,T)\sigma_1(T)}{\sigma_0^2(T)}=\frac{J}{\pi}\int_0^{2\pi}dk{\rm e}^{-\beta E_{magn}(k)}(\cos{k}+\cos{q}).
\end{eqnarray}
Using a shift of the integration variable one readily gets from (B.3) and (B.8)
\begin{equation}
\theta_1(\omega,q,T)+\sigma_1(T)(\omega-E_{magn}(q))=\frac{1}{\pi}\int_{-\pi}^{\pi}dk{\rm e}^{-\beta E_{magn}(k-q)}(h-\omega+J\cos{(k-q)}).
\end{equation}

\section{Evaluation of the boundary frequencies}
\setcounter{equation}{0}

\subsection{Foundations}

According to (105) at $k=\pm\pi$ the function $\Phi_{down}(q,k)$ ($\Phi_{up}(q,k)$) takes its maximum (minimum) value which in fact is $E_{magn}(q)$.
Hence for fixed $q$ and $k\in[-\pi,\pi]$ the function $\Phi_{down}(q,k)$ ($\Phi_{up}(q,k)$) should take its minimum (maximum) values namely $\omega_{min}(q)$ ($\omega_{max}(q)$)
only at local extremum points $k=k_{min}(q)$ and $k=k_{max}(q)$. In other words
\begin{equation}
\omega_{min}(q)=\Phi_{down}(q,k_{min}(q)),\qquad\omega_{max}(q)=\Phi_{up}(q,k_{max}(q)).
\end{equation}
Using the short notations $k_{min}$ and $k_{max}$ instead of $k_{min}(q)$ and $k_{max}(q)$ one readily gets
\begin{eqnarray}
&&\frac{\partial\Phi_{down}(q,k)}{\partial k}\Big|_{k=k_{min}}=0\Longleftrightarrow\sin{(k_{min}(q)-q-k_{gap})}=\sin{\frac{k_{min}}{2}},\\
&&\frac{\partial^2\Phi_{down}(q,k)}{\partial k^2}\Big|_{k=k_{min}}>0\Longleftrightarrow2\cos{(k_{min}-q-k_{gap})}-\cos{\frac{k_{min}}{2}}<0,\\
&&\frac{\partial\Phi_{up}(q,k)}{\partial k}\Big|_{k=k_{max}}=0\Longleftrightarrow\sin{(k_{max}-q-k_{gap})}=-\sin{\frac{k_{max}}{2}},\\
&&\frac{\partial^2\Phi_{up}(q,k)}{\partial k^2}\Big|_{k=k_{max}}<0\Longleftrightarrow2\cos{(k_{max}-q-k_{gap})}+\cos{\frac{k_{max}}{2}}>0.
\end{eqnarray}
From (C.2) and (C.4) follows that
\begin{equation}
\Big|\cos{(k_{min}-q-k_{gap})}\Big|=\Big|\cos{\frac{k_{min}}{2}}\Big|,\qquad\Big|\cos{(k_{max}-q-k_{gap})}\Big|=\Big|\cos{\frac{k_{max}}{2}}\Big|.
\end{equation}
At the same time according to (39) $\cos{k_{min}/2}\geq0$ and $\cos{k_{max}/2}\geq0$ and hence in agreement with (C.3) and (C.5) one has from (C.6)
\begin{equation}
\cos{(k_{min}-q-k_{gap})}=-\cos{\frac{k_{min}}{2}},\qquad\cos{(k_{max}-q-k_{gap})}=\cos{\frac{k_{max}}{2}}.
\end{equation}
Equations (100), (C.1) and (C.7) yield
\begin{equation}
\omega_{min}(q)=h-3|J|\cos{\frac{k_{min}(q)}{2}},\qquad\omega_{max}(q)=h+3|J|\cos{\frac{k_{max}(q)}{2}}.
\end{equation}

According to (C.2) and (C.7)
\begin{eqnarray}
&&\sin{\Big(\frac{3k_{min}}{2}-q-k_{gap}\Big)}=\sin{(k_{min}-q-k_{gap})}\cos{\frac{k_{min}}{2}}\nonumber\\
&&+\cos{(k_{min}-q-k_{gap})}\sin{\frac{k_{min}}{2}}=0,\nonumber\\
&&\cos{\Big(\frac{3k_{min}}{2}-q-k_{gap}\Big)}=\cos{(k_{min}-q-k_{gap})}\cos{\frac{k_{min}}{2}}\nonumber\\
&&-\sin{(k_{min}-q-k_{gap})}\sin{\frac{k_{min}}{2}}=-\Big(\cos^2{\frac{k_{min}}{2}}+\sin^2{\frac{k_{min}}{2}}\Big)=-1.
\end{eqnarray}
In the same manner (C.4) and (C.7) yield
\begin{equation}
\sin{\Big(\frac{3k_{max}}{2}-q-k_{gap}\Big)}=0,\qquad\cos{\Big(\frac{3k_{max}}{2}-q-k_{gap}\Big)}=1.
\end{equation}

According to (C.9) and (C.10) one has
\begin{equation}
k_{min}(q)=\frac{2}{3}\Big(q+k_{gap}+\pi\Big)+\frac{4j_{min}\pi}{3},\qquad
k_{max}(q)=\frac{2}{3}\Big(q+k_{gap}\Big)+\frac{4j_{max}\pi}{3},
\end{equation}
where the integers $j_{min}$ and $j_{max}$ should ensure the condition (39).

\subsection{Minimum at $k_{gap}=0$}

For $k_{gap}=0$ the integer $j_{min}$ in (C.11) takes only two values which agree with (39)
\begin{eqnarray}
&&j_{min,1}=0:\qquad q\in\Big[-\pi,\frac{\pi}{2}\Big),\quad k_{min,1}=\frac{2(q+\pi)}{3},\nonumber\\
&&j_{min,2}=-1:\qquad q\in\Big(-\frac{\pi}{2},\pi\Big],\quad k_{min,2}=\frac{2(q-\pi)}{3},
\end{eqnarray}
(we have excluded the two boundary points $q=\pm\pi/2$ related to $k_{min}=\pm\pi$ for which the inequality (C.3) turns into an equality and the minima turn into inflection points).
As it follows from (C.12) at $k\in(-\pi/2,\pi/2)$ there is a pair of solutions related to two local minima. So additionally to the principal minimum
\begin{eqnarray}
&&k_{min}(q)=\frac{2(q+\pi)}{3},\qquad q\in[-\pi,0],\nonumber\\
&&k_{min}(q)=\frac{2(q-\pi)}{3},\qquad q\in[0,\pi],
\end{eqnarray}
for which
\begin{equation}
\omega_{min}(q)=\Phi_{down}(q,k_{min}(q))=h-3|J|\cos{\frac{\pi-|q|}{3}},
\end{equation}
there is an additional one related to
\begin{eqnarray}
&&k_c(q)=\frac{2(q-\pi)}{3},\qquad q\in(-\pi/2,0],\nonumber\\
&&k_c(q)=\frac{2(q+\pi)}{3},\qquad q\in[0,\pi/2),
\end{eqnarray}
and for which
\begin{equation}
\omega_c(q)=\Phi_{down}(q,k_c(q))=h-3|J|\cos{\frac{\pi+|q|}{3}},\qquad0\leq|q|<\frac{\pi}{2}.
\end{equation}
According to (C.14) and (C.16)
\begin{equation}
\omega_{min}(q)<\omega_c(q),\quad q\neq0,
\end{equation}
but
\begin{equation}
\omega_{min}(0)=\omega_c(0)=h-3|J|\cos{\frac{\pi}{3}}=h-\frac{3|J|}{2}.
\end{equation}
Hence at $q=k_{gap}=0$ the function $\Phi_{down}(0,k)$ has two equal local minima in the points
\begin{equation}
k_{\pm}=\pm\frac{2\pi}{3}.
\end{equation}

\subsection{Maximum at $k_{gap}=0$}
In a similar manner for $k_{gap}=0$ the integer $j_{max}$ in (C.11) takes only three values which agree with (39). Namely
\begin{eqnarray}
&&j_{max,1}=1:\quad q\in\Big[-\pi,-\frac{\pi}{2}\Big),\quad k_{max,1}=\frac{2(q+2\pi)}{3},\nonumber\\
&&j_{max,2}=0:\quad q\in\Big[-\pi,\pi\Big],\quad k_{max,2}=\frac{2q}{3},\nonumber\\
&&j_{max,3}=-1:\quad q\in\Big(\frac{\pi}{2},\pi\Big],\quad k_{max,3}=\frac{2(q-2\pi)}{3},
\end{eqnarray}
(we have excluded the two boundary points $q=\pm\pi/2$ related to $k_{max}=\pm\pi$ for which the inequality (C.5) turns into an equality and the maxima turn into inflection points).
Substituting (C.20) into (C.1) we conclude that the principal and additional maxima correspond to
\begin{eqnarray}
&&k_{max}(q)=\frac{2q}{3},\qquad q\in[-\pi,\pi],\\
&&k_s(q)=\frac{2(q+2\pi)}{3},\qquad q\in[-\pi,-\pi/2),\nonumber\\
&&k_s(q)=\frac{2(q-2\pi)}{3},\qquad q\in(\pi/2,\pi],
\end{eqnarray}
so that (at $k_{gap}=0$)
\begin{eqnarray}
&&\omega_{max}(q)=\Phi_{up}(q,k_{max}(q))=h+3|J|\cos{\frac{q}{3}}\qquad-\pi<q\leq\pi,\\
&&\omega_s(q)=\Phi_{up}(q,k_s(q))=h+3|J|\cos{\frac{2\pi-|q|}{3}},\qquad\frac{\pi}{2}<|q|\leq\pi.
\end{eqnarray}

According to (C.23) and (C.24)
\begin{equation}
\omega_s(q)<\omega_{max}(q),\quad q\neq\pm\pi,
\end{equation}
but
\begin{equation}
\omega_{max}(\pi)=\omega_s(\pi)=h+3|J|\cos{\frac{\pi}{3}}=h+\frac{3|J|}{2}.
\end{equation}
Hence at $q=\pi-k_{gap}=\pi$ the function $\Phi_{up}(\pi,k)$ has two equal local maxima at the points (C.19).

\subsection{Minimum and maximum at $k_{gap}=\pi$}

Let us include $k_{gap}$ in the notations (100) writing $\Phi_{down}(q,k,k_{gap})$ and $\Phi_{up}(q,k,k_{gap})$ instead of $\Phi_{down}(q,k)$ and $\Phi_{up}(q,k)$. Then according to (100)
\begin{equation}
\Phi_{down}(k,q,\pi)=2h-\Phi_{up}(k,q,0),\qquad\Phi_{up}(k,q,\pi)=2h-\Phi_{down}(k,q,0).
\end{equation}
Hence minima of $\Phi_{down}(k,q,\pi)$ and maxima of $\Phi_{up}(k,q,\pi)$ are in one to one correspondence with maxima of $\Phi_{up}(k,q,0)$ and minima of $\Phi_{down}(k,q,0)$.
Using now the results of the previous two subsections one readily gets for $k_{gap}=\pi$ ($J<0$) the following list of relations
\begin{eqnarray}
&&k_{min}=\frac{2q}{3},\qquad q\in[-\pi,\pi],\\
&&k_c=\frac{2(q+2\pi)}{3},\qquad q\in[-\pi,-\pi/2),\nonumber\\
&&k_c=\frac{2(q-2\pi)}{3},\qquad q\in(\pi/2,\pi],\\
&&k_{max}=\frac{2(q+\pi)}{3},\qquad q\in[-\pi,0],\nonumber\\
&&k_{max}=\frac{2(q-\pi)}{3},\qquad q\in[0,\pi],\\
&&k_s=\frac{2(q-\pi)}{3},\qquad q\in(-\pi/2,0],\nonumber\\
&&k_s=\frac{2(q+\pi)}{3},\qquad q\in[0,\pi/2),\\
&&\omega_{min}(q)=h-3|J|\cos{\frac{q}{3}},\\
&&\omega_c(q)=h-3|J|\cos{\frac{2\pi-|q|}{3}},\\
&&\omega_{max}(q)=h+3|J|\cos{\frac{\pi-|q|}{3}},\\
&&\omega_s(q)=h+3|J|\cos{\frac{\pi+|q|}{3}}.
\end{eqnarray}

Also
\begin{equation}
\omega_{min}(q)<\omega_c(q),\qquad q\neq\pm\pi,\qquad\omega_s(q)<\omega_{max}(q),\quad q\neq0,
\end{equation}
and
\begin{equation}
\omega_c(\pm\pi)=\omega_{min}(\pm\pi)=h-\frac{3|J|}{2},\qquad\omega_s(0)=\omega_{max}(0)=h+\frac{3|J|}{2}.
\end{equation}

\section{Solutions of the quartic equation}
\setcounter{equation}{0}

According to the identity
\begin{equation}
(x^2-1)(x-1)^2+\frac{t^3}{4(t+1)}=\Big(x^2-x+\frac{t}{2}\Big)^2-(t+1)\Big(x-\frac{t+2}{2(t+1)}\Big)^2,
\end{equation}
the quartic equation
\begin{equation}
(x^2-1)(x-1)^2+\frac{t^3}{4(t+1)}=0,
\end{equation}
splits on the pair of the following quadratic ones
\begin{eqnarray}
&&x^2-(1+\sqrt{t+1})x+\frac{1}{2}\Big(t+\frac{t+2}{\sqrt{t+1}}\Big)=0,\\
&&x^2-(1-\sqrt{t+1})x+\frac{1}{2}\Big(t-\frac{t+2}{\sqrt{t+1}}\Big)=0.
\end{eqnarray}
Hence in order to solve (154) we have at first solve the cubic equation
\begin{equation}
t^3-4\lambda_0^2(t+1)=0,\qquad0\leq4\lambda_0^2\leq\frac{27}{4}.
\end{equation}

Using the Tartaglia substitution
\begin{equation}
t=u_++u_-,\qquad3u_+u_-=4\lambda_0^2,
\end{equation}
we readily get from (D.5) and (D.6)
\begin{equation}
u_+^3+u_-^3=4\lambda_0^2,\qquad u_+^3u_-^3=\frac{64\lambda_0^6}{27}.
\end{equation}
Hence the pair $u_{\pm}^3$ is the pair of solutions of the quadratic equation
\begin{equation}
z^2-4\lambda_0^2z+\frac{64\lambda_0^6}{27}=0.
\end{equation}
Namely
\begin{equation}
u_{\pm}^3=2\lambda_0^2\Big(1\pm\sqrt{1-\frac{16\lambda_0^2}{27}}\Big).
\end{equation}
From (D.6) and (D.9) follows that at $16\lambda_0^2\neq27$ (158) is the single {\it real} solution of (D.5).

Turning to the quadratic equations (D.3) and (D.4) we readily calculate their discriminants
\begin{eqnarray}
&&D_1(t)=2-t-\frac{2}{\sqrt{t+1}},\\
&&D_2(t)=2-t+\frac{2}{\sqrt{t+1}}.
\end{eqnarray}
It may be readily seen that $D_1(t)<0$ for all $t\geq-1$ except the point $t=0$ where $D_1(0)=0$ and (D.3) has the two-fold solution $x=1$ (since in this case $x=1$ is the three-fold solution of (163)
this two solutions are in fact the extra ones). At the same time $D_2(t)\geq0$ on the whole interval $-1<t\leq3$. Hence the pair of real solutions of (D.1) should be obtained from (D.4) and hence
has the form (157).

\begin{figure}
\includegraphics[width=10cm]{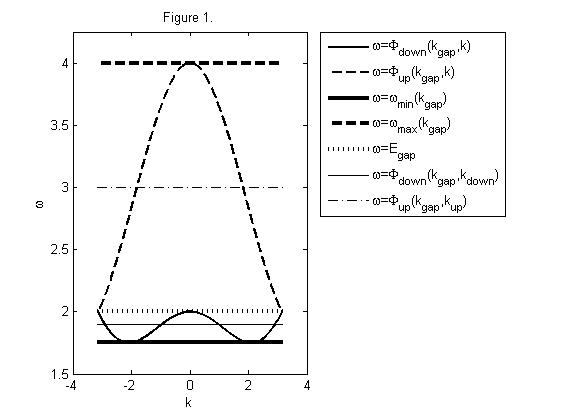}
\end{figure}
\begin{figure}
\includegraphics[width=10cm]{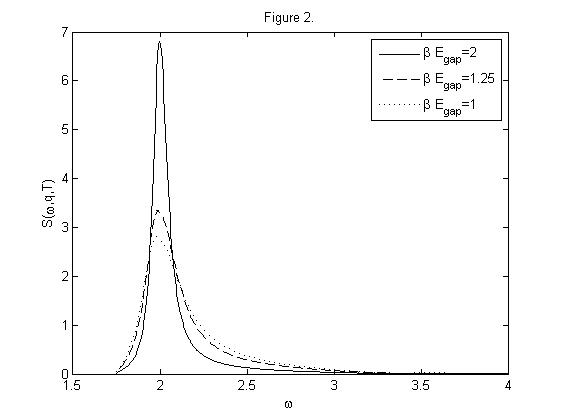}
\end{figure}
\begin{figure}
\includegraphics[width=10cm]{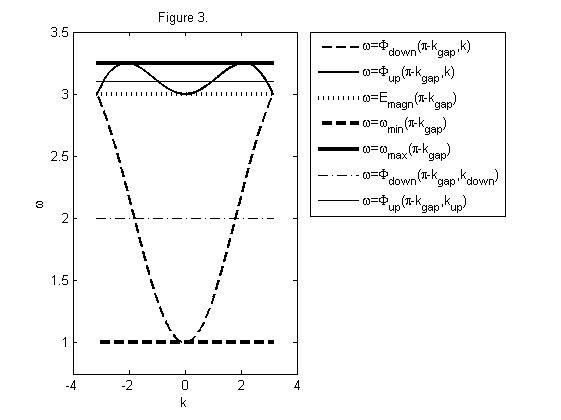}
\end{figure}
\begin{figure}
\includegraphics[width=10cm]{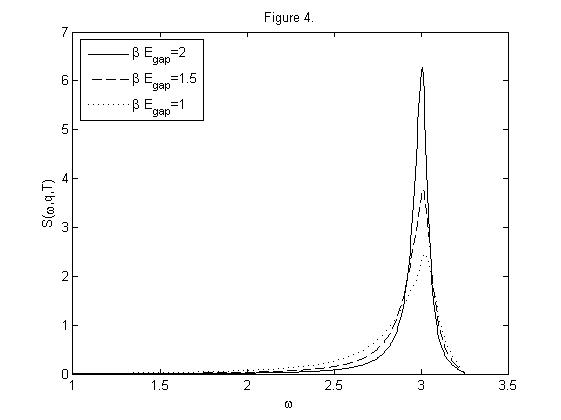}
\end{figure}
\begin{figure}
\includegraphics[width=10cm]{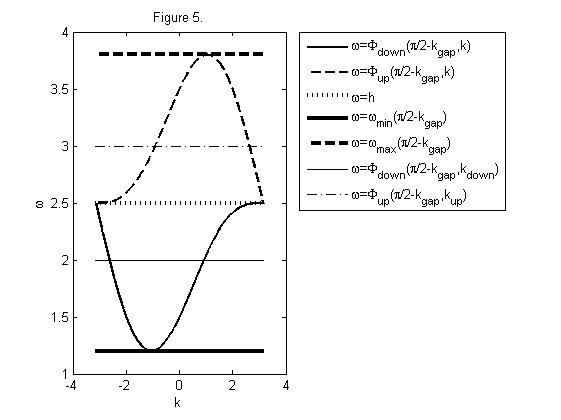}
\end{figure}
\begin{figure}
\includegraphics[width=10cm]{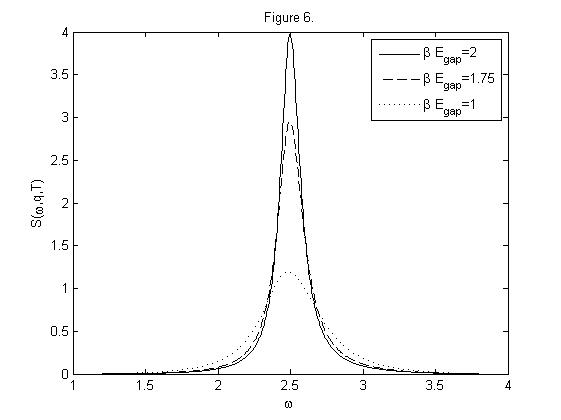}
\end{figure}

\end{document}